# Physical Layer Network Coding Schemes over Finite and Infinite Fields

Shengli Zhang, Soung Chang Liew, Lu Lu
The Department of Information Engineering
The Chinese University of Hong Kong, Hong Kong, China
Email: {slzhang5, soung, ll007}@ie.cuhk.edu.hk

*Abstract:* **Direct application of network coding at the physical layer - physical layer network coding (PNC) - is a promising technique for two-way relay wireless networks. In a two-way relay network, relay nodes are used to relay two-way information flows between pairs of end nodes. This paper proposes a precise definition for PNC. Specifically, in PNC, a relay node does not decode the source information from the two ends separately, but rather directly maps the combined signals received simultaneously to a signal to be relayed. Based on this definition, PNC can be further sub-classed into two categories - PNCF (PNC over finite field) and PNCI (PNC over infinite field) - according to whether the network-code field (or groups, rings) adopted is finite or infinite. For each of PNCF and PNCI, we consider two specific estimation techniques for dealing with noise in the mapping process. The performance of the four schemes is investigated by means of analysis and simulation, assuming symbol-level synchronization only.**

## I. INTRODUCTION

Network coding, first proposed in [1], is a promising technique for achieving max-flow min-cut capacity in multicast transmission. Linear network coding has been shown to be powerful enough to achieve this capacity [2]. This paper focuses on the use of linear network coding in wireless networks. Although the original investigation of network coding was in the context of wired networks, its potential to boost performance in wireless networks could be even more significant thanks to the broadcast nature of the wireless medium [3]. Ref. [3] shows that direct application of network coding at the physical layer in a wireless relay network could double the capacity of bidirectional point-to-point communication. A similar idea was independently presented in [4]. The analog network coding proposed in [5] is essentially another variation of physical layer network coding (PNC).

Several other PNC schemes have also been proposed for the wireless two-way relay channel (TWRC). For example, [6] proposed a PNC scheme based on Tomlinson-Harashima precoding. In [7], a number of memoryless relay functions, including the BER optimal function, were identified and analyzed assuming phase synchronization between signals of the transmitters. Under the general definition for PNC given in our paper here, there is a one-to-one correspondence between a relay function and a specific PNC scheme.

We give a precise definition for PNC to distinguish it from the traditional straightforward network coding (SNC). Additionally, we classify PNC schemes into two categories - PNCF (PNC over finite field) and PNCI (PNC over infinite field) - according to whether the network-code field (or groups, rings) adopted is finite or infinite.

With the definition and classification, the construction of a PNC scheme can be regarded as consisting of two parts: (i) determination of the network code to be used at the relay node; (ii) computation of the information to be relayed at the relay node based on the signals received from the two end nodes. In this paper, we investigate several well-known signal estimation techniques for (ii) under PNCF and PNCI. Among the resulting schemes, one is ANC, one is a novel scheme and the other two are the generalizations of the PNC schemes in [3, 7]. For all the four PNC schemes, only symbol-level synchronization is assumed.

## II. SYSTEM MODEL AND DEFINITIONS

### A. System model

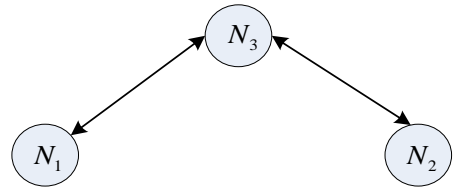

Figure 1. Two way relay channel

We consider the two-way relay channel as shown in Fig.1, in which nodes $N_1$ and $N_2$ exchange information with the help of relay node $N_3$. We assume that all nodes are half-duplex, i.e., a node can not receive and transmit simultaneously. This is an assumption arising from practical considerations because it is difficult for the wireless nodes to remove the strong interference of its own transmitting signal from the received signal. We also assume that there is no direct link between $N_1$ and $N_2$. An example in practice is a satellite communication system in which the two end nodes on the earth can only communicate with each other via the relay satellite. In this paper, $W_i$ denotes the un-coded packet of $N_i$; $X_i$ denotes the corresponding transmitted packet after channel coding and modulation; and $Y_i$ denotes the received base-band packet at $N_i$. A lowercase letter, *w*, *x*, or *y*, denotes one symbol within the corresponding packet.

Because of half-duplexity and the lack of a direct link between $N_1$ and $N_2$, the transmission must consist two phases: the uplink and downlink phases. Under PNC, for the uplink phase, $N_1$ and $N_2$ transmit to $N_3$ at the same time. Therefore, $N_3$ receives

$$y_3 = h_{13}x_1 + h_{23}x_2 + n_3 \qquad (1)$$

where $n_i$ is the noise at $N_i$ assumed to be complex Gaussian with unit variance; and $h_{ij}$ is the complex path-loss coefficient for the channel from $N_i$ to $N_j$. Fading is not considered in this paper. For all the packets ($X_1$, $X_2$ and $X_3$), QPSK modulation is assumed, and the transmitting power is normalized to 2, which is the QPSK signal's variance. In (1), symbol-level synchronization is implied. The carrier phase offset and the transmitting power differences are combined into $h_{ij}$. We also assume that the path-loss coefficients $h_{ij}$ can be perfectly estimated by each receiving node $N_i$. Furthermore, the relay node $N_3$ will forward $h_{i3}$ estimated by to both end nodes.

In the downlink phase, $N_3$ generates a new signal, $x_3$, based on the received signal $y_3$, and broadcasts it to both $N_1$ and $N_2$. We can write the signals received by $N_1$ and $N_2$ as

$$y_1 = h_{31}x_3 + n_1 \quad y_2 = h_{32}x_3 + n_2 \quad (2)$$

The target information to be received at the destinations, $x_1$ at $N_2$ and $x_2$ at $N_1$, will be decoded from $y_1$ and $y_2$, respectively, with the help of the nodes' self-information. In general, $x_3$ must be a function of $x_1$ and $x_2$, denoted by $x_3 = f(x_1, x_2)$. The following sub-section defines and classifies PNC based on different forms of the function $f$.

### B. Definition and classification of PNC

For comparison purposes, let us review SNC briefly. Traditionally, SNC [8] is regarded as an upper layer technique and is separated from other lower-layer signal processes such as modulation and channel coding. It works as follows. $N_3$ first decodes $W_1$ and $W_2$ separately. Then it encodes $W_1$ and $W_2$ into a network coded version $W_3$. After that, $W_3$ is channel encoded and modulated into $X_3$ before being sent out

PNC was inspired by the observation that it is unnecessary for the relay node to know the exact source information [3]. With PNC, $N_3$ transforms $Y_3$ directly to a network coded version of the combined input symbols without detecting the individual input symbols separately. More formally, PNC is defined as follows:

***Definition 1:*** *Physical layer network coding is the coding operation which directly transforms the received baseband signal $y_3$ in (1) to a network-coded symbol $x_3 = f(x_1, x_2)$ for relay, without separate detection of $x_1$ and $x_2$.*

If the network code is over a finite field (e.g., $GF(2)$) then the PNC scheme belongs to PNCF. If there were no noise, the transformation from $y_3$ to $x_3 = x_1 \oplus x_2$ [1] would be deterministic. However, due to noise, PNCF generates an estimation of $x_1 \oplus x_2$ from the received signal $y_3$.

If the network code is over an infinite field (e.g., the real field $R$ or complex field $C$ [10]) then the PNC scheme belongs to PNCI. In this paper, the network code of $x_1$ and $x_2$ over complex field is fixed to $x_3 = h_{13}x_1 + h_{23}x_2$ to match the multiple access channels, as in [5, 10]. Without noise, the received signal, $y_3 = h_{13}x_1 + h_{23}x_2$, is already in the form of real/complex field network coding. To deal with noise, PNCI generates an estimation of $h_{13}x_1 + h_{23}x_2$ from $y_3$.

### III. PARTICULAR PNC SCHEMES

The previous section gives a general definition for PNC, and divides PNC schemes into two broad subclasses - PNCF and PNCI – according to the network code used. Given a network code, we could also use different estimation techniques at the relay node to compute the relayed value. This section considers three specific estimation functions.

### A. PNCF schemes

Under PNCF, we want to estimate $x_1 \oplus x_2$ from $y_3$. We introduce two estimation methods here.

### A.1 MAP-based PNCF

The estimator for maximum a posteriori (MAP) estimation is

$$x_3 = \alpha \arg \max_{x_1 \oplus x_2 \in \{\pm 1 \pm j\}} \Pr(y_3 | x_1 \oplus x_2) P(x_1 \oplus x_2) \quad (3)$$

where $\alpha$ is a coefficient to constrain the average power of $x_3$, and $P(x_1 \oplus x_2)$ is the a priori distribution of $x_1 \oplus x_2$, which is a constant of 1/4 under QPSK modulation. We can rewrite (3) as

$$(a,b,c,d) = \arg \max_{a',b',c',d' \in \{\pm 1\}}$$
$$\{P_{a'b'c'd'} + P_{(-a')b'(-c')d'} + P_{a'(-b')c'(-d')} + P_{(-a')(-b')(-c')(-d')}\} \quad (4)$$
$$x_3 = a \oplus c + j(b \oplus d)$$

where $j = \sqrt{-1}$ and $P_{abcd}$, $a,b,c,d \in \{\pm 1\}$, is the probability that $y_3$ is received when $x_1 = a+jb$ and $x_2 = c+jd$, given by

$$\begin{aligned}P_{abcd} &= \Pr(y_3 | x_1 = a+jb, x_2 = c+jd) \\ &= \varphi(y_3 - h_{13}(a+jb) - h_{23}(c+jd), 1)\end{aligned} \quad (5)$$

where $\varphi(\mu, \sigma^2) = \frac{1}{2\pi\sigma^2} e^{-|\mu|^2/2\sigma^2}$.

Only symbol-level synchronization is assumed in the MAP-based PNCF in (4). If we have the stronger condition of power and carrier-phase synchronization such that $h_{13} = h_{23} = h_0$ is a real value, then (4) becomes

$$\begin{aligned}x_3 &= sign(\varphi(y_3^R - 2h_0, 1) + \varphi(y_3^R + 2h_0, 1) - 2\varphi(y_3^R, 1)) \\ &+ j sign(\varphi(y_3^I - 2h_0, 1) + \varphi(y_3^I + 2h_0, 1) - 2\varphi(y_3^I, 1))\end{aligned} \quad (6)$$

where $y_3^R$ is the real part of the received symbol and $y_3^I$ is the imaginary part. It is easy to verify that (6) is equivalent to the original PNC scheme introduced in [3]. We could regard (4) as a generalization of the original PNC scheme.

---

[1] For QPSK, $x_1, x_2 \in \{\pm 1 \pm j\}$, $x_1 \oplus x_2 = \text{Re}(x_1) \cdot \text{Re}(x_2) + j \text{Im}(x_1) \cdot \text{Im}(x_2)$. This definition is equivalent to the GF(2) addition of $x_1, x_2$ for the real and imaginary parts separately.

## A. 2MMSE-based PNCF

Another estimation method is the minimum mean square error (MMSE) estimation. Applying MMSE estimation at the relay node, we can obtain

$$x_3 = \alpha \varepsilon\{x_1 \oplus x_2 | y_3\} = \alpha \sum_{t \in \{\pm 1 \pm j\}} t \Pr(t | y_3) \quad (7)$$

where $\alpha$ has the same meaning as in (3) and $\varepsilon\{x_1 \oplus x_2 | y_3\}$ is the MMSE estimation of $x_1 \oplus x_2$ given $y_3$. We have

$$\varepsilon\{x_1 \oplus x_2 | y_3\} = \sum_{t \in \{\pm 1 \pm j\}} t \Pr(t | y_3) = \sum_{t \in \{\pm 1 \pm j\}} \frac{t \Pr(y_3 | t) \Pr(t)}{\sum_{s \in \{\pm 1 \pm j\}} \Pr(y_3 | s)}$$

$$= \frac{1}{4} \sum_{a,b,c,d \in \{\pm 1\}} \frac{(a \oplus c + jb \oplus d) P_{abcd}}{\sum_{a',b',c',d' \in \{\pm 1\}} P_{a'b'c'd'}} \quad (8)$$

The MMSE estimation of the whole packet, $\varepsilon\{X_1 \oplus X_2 | Y_3\}$, can be obtained by estimating every symbol in the packet as in (8). Writing the power constraint $\alpha = \sqrt{2/E|\varepsilon\{X_1 \oplus X_2 | Y_3\}|^2}$ explicitly (recall that $\alpha$ constrains the average power of the overall packet, and $E(|\varepsilon\{X_1 \oplus X_2 | Y_3\}|^2)$ is the average power of the estimated packet), we can substitute (8) into (7) to obtain

$$x_3 = \sqrt{\frac{2}{E|\varepsilon\{X_1 \oplus X_2 | Y_3\}|^2}} \sum_{a,b,c,d \in \{\pm 1\}} \frac{(a \oplus c + jb \oplus d) P_{abcd}}{4 \sum_{a',b',c',d' \in \{\pm 1\}} P_{a'b'c'd'}} \quad (9)$$

Note that the $x_3$ estimated in (9) is a continuous value, while the $x_3$ estimated in (4) is discrete. With the stronger power and carrier-phase synchronization assumption, i.e. $h_{13} = h_{23} = h_0$ is a real value, (8) can be rewritten as

$$\frac{1}{2} \frac{\cosh 2h_0 y_3^R - \exp(2h_0^2)}{\cosh 2h_0 y_3^R + \exp(2h_0^2)} + j\frac{1}{2} \frac{\cosh 2h_0 y_3^I - \exp(2h_0^2)}{\cosh 2h_0 y_3^I + \exp(2h_0^2)} \quad (10)$$

Eqn. (10) is equivalent to the eqn. (22) in [7], which describes an estimate-and-forward scheme. The threshold in (22) in [7], however, is not explicitly specified. Eqn. (10), on the other hand, makes use of the optimal threshold value. Furthermore, unlike (22) om [7], we have not assumed carrier-phase synchronization in (9). The MMSE-based PNCF is more general than the scheme in [7] in that it encompasses both cases with and without carrier-phase synchronization. The case without carrier-phase synchronization is of practical interest because of its easier implementation.

## B. PNCI schemes

We now consider the PNCI schemes, whose objective is to estimate $h_{13}x_1 + h_{23}x_2$ from $y_3$.

## B.1 Linear-MMSE-based PNCI

Linear MMSE estimation is widely used due to its simplicity and good performance. If we use linear MMSE to estimate $h_{13}x_1 + h_{23}x_2$, we get

$$x_3 = \alpha y_3 = \sqrt{2/(2(|h_{13}|^2 + |h_{23}|^2) + 1)} y_3 \quad (11)$$

where $\alpha$ is the power constraint coefficient. The definition of Linear-MMSE-based PNCI scheme in (11) is simply the ANC scheme of [5].

We can also see from (11) that the linear MMSE-based PNCI scheme only depends on the absolute value of the channel coefficients. As a result, unlike in the PNCF schemes, the carrier phase offset between the signals of the end nodes will not affect the performance here.

## B.2 MMSE-based PNCI

According to estimation theory, the linear MMSE estimation is optimal in terms of minimizing the estimation error when the distribution of $h_{13}x_1 + h_{23}x_2$ is Gaussian. For non-Gaussian distribution as in our system (due to the assumption of QPSK modulation at $N_1$ and $N_2$), the unconstrained MMSE estimation should perform better. The general MMSE-based PNCI scheme leads to

$$x_3 = \alpha \sum_{t_1 \in \{\pm 1 \pm j\}} \sum_{t_2 \in \{\pm 1 \pm j\}} (h_{13}t_1 + h_{23}t_2) \Pr(x_1 = t_1, x_2 = t_2 | y_3) \quad (12)$$

With similar derivation and transformation as before, we can rewrite (12) as

$$x_3 = \alpha' \sum_{a,b,c,d \in \{\pm 1\}} \frac{[h_{13}(a+jb) + h_{23}(c+jd)] P_{abcd}}{\sum_{a',b',c',d' \in \{\pm 1\}} P_{a'b'c'd'}} \quad (13)$$

where $\alpha = \frac{1}{16}\alpha' = \frac{1}{16}\sqrt{2/E(|\varepsilon\{X_3|Y_3\}|^2)}$ are coefficients to satisfy the power constraint.

With the stronger power and carrier-phase synchronization assumption, i.e. $h_{13} = h_{23} = h_0$ is a real value, (12) can be rewritten as

$$x_3 = 2\alpha' \sum_{a,b \in \{1,0,-1\}} \frac{(a+bj)\varphi(y_3^R - 2ah_0, 1)\varphi(y_3^I - 2bh_0, 1)}{\sum_{a',b' \in \{1,0,-1\}} \varphi(y_3^R - 2a'h_0, 1)\varphi(y_3^I - 2b'h_0, 1)} \quad (14)$$

## IV. PERFORMANCE ANALYSIS AND SIMULATION

In this section, we first review the concept of the generalized SNR (GSNR), originally proposed in [9] as a measure for the quality of memoryless relay channels. Then we analyze the GSNR performance of the four PNC schemes in the previous section. We prove that MMSE estimation maximizes the GSNR at the destination nodes. In addition, we compare the GSNR and BER performance of the four schemes based on numerical simulation.

## A. Review of GSNR

GSNR was originally proposed [9] as a measure of the quality of *one-way* relay channels. Consider a three-node one-way relay system, where $x$ is the signal from the source,

$r = g(x+n_1)$ is the regenerated signal at the relay, and $y = r+n_2$ is the received signal at the destination. If $g(x+n_1) = \alpha(x+n_1)$ is a linear function, then $y = \alpha(x+n_1+n_2/\alpha)$. The SNR at the destination is

$$SNR = \frac{P_x}{P_n} = \frac{\alpha^2 E[|x|^2]}{\alpha^2 E[|n_1|^2] + E[|n_1|^2]} \quad (15)$$

where $P_x$ is the power of the data $x$ and $P_n$ is the power of the noise. If $g$ is a non-linear function, however, the signal $y$ can not be written as the addition of $x$ and an independent noise directly. And the definition of SNR is not so clear as (15). To deal with regenerate function $g$ that could be nonlinear in general, we could express $y$ in the following form [9]:

$$y = \frac{E[x^*y]}{E[|x|^2]}(x + e_u) \quad (16)$$

where superscript $*$ denotes the conjugate transpose. It is easy to verify that the error $e_u = \frac{E[|x|^2]}{E[x^*y]}y - x$ is uncorrelated to $x$. The GSNR is defined as [9]

$$GSNR = \frac{P_x}{MSUE} = \frac{E[|x|^2]}{E[|e_u|^2]} \quad (17)$$

where *MSUE* denotes the mean square uncorrelated error (uncorrelated to data $x$).

Simulations in [9] showed that there is a correspondence between GSNR and BER when BPSK modulation is assumed (larger GSNR implies smaller BER, and vice versa). GSNR, however, is easier to analyze than BER for general relay (regeneration) functions, and is a more convenient metric for analysis. For PNC schemes studied here, the numerical simulation in the second sub-section also shows that there is a correspondence between GSNR and BER in our systems.

### B. GSNR Analysis

This sub-section gives the GSNR expressions of the four PNC schemes.

**MAP-based PNCF**

First, we consider MAP-based PNCF. According to the expression in (16), we rewrite the relay function as

$$x_3 = x_1 \oplus x_2 + d = \frac{E((x_1 \oplus x_2)^* x_3)}{2}(x_1 \oplus x_2 + e_u) \quad (18)$$

According to (4), the Euclidean distance between $x_3$ and the original symbol $x_1 \oplus x_2$ is

$$d = \begin{cases} 0 & 1-\varepsilon_0-\varepsilon_1-\varepsilon_2 \\ (x_1 \oplus x_2)^* - x_1 \oplus x_2 & \varepsilon_0 \\ -(x_1 \oplus x_2)^* - x_1 \oplus x_2 & \varepsilon_1 \\ -2(x_1 \oplus x_2) & \varepsilon_2 \end{cases} \quad (19)$$

where $\varepsilon_0$ ($\varepsilon_1$) is the probability that only the imaginary (real) part of the estimation is wrong, and $\varepsilon_2$ is the probability that both the real and imaginary parts of $x_3$ do not equal to those of $x_1 \oplus x_2$. The probability of $\varepsilon_0$, $\varepsilon_1$, and $\varepsilon_2$ can be calculated from (4). Then we can calculate the power constraint coefficient

$$E(\frac{(x_1 \oplus x_2)^* x_3}{2}) = 1 + \frac{E((x_1 \oplus x_2)^* d)}{2} = 1 - (\varepsilon_0 + \varepsilon_1 + 2\varepsilon_2) \quad (20)$$

Based on (20), we can calculate $\varepsilon_u$ in (18) and obtain the MSUE at the relay

$$MSUE_{PNCF\_MAP} = E(|e_u|^2) = 2/(1-\varepsilon_0-\varepsilon_1-\varepsilon_2)^2 - 2 \quad (21)$$

The GSNR at the destination node $N_i$, denoted by $GSNR_i$, can be obtained based on (21). *Appendix 1* shows that

$$GSNR_i = \frac{2\beta^2 |h_{3i}|^2}{\beta^2 |h_{3i}|^2 MSUE + 1} \quad i = 1,2 \quad (22)$$

where $\beta = 2/\sqrt{2+MSUE}$.

**MMSE-based PNCF**

For MMSE-based PNCF, the MSUE at the relay node is

$$MSUE_{PNCF\_MMSE} = 2(2\text{Re}(\lambda)+1) - |\lambda^*+1|^2 E_{y_3}[|\varepsilon(x_1 \oplus x_2 | y_3)|^2] \quad (23)$$

where $\lambda^* = 2/E[(x_1 \oplus x_2)^* \varepsilon(x_1 \oplus x_2 | y_3)] - 1$. The derivation of (23), omitted here, is similar to the proof of *Theorem 1* in [9]. The GSNR at the end nodes can be obtained by substituting (23) into (22).

**Linear MMSE-based PNCI**

For linear MMSE-based PNCI, the MSUE analysis is simple because the uncorrelated error is just the Gaussian noise which is assumed to be normalized. Therefore, the MSUE at the relay node is

$$MSUE_{PNCI\_linearMMSE} = 2 \quad (24)$$

The GSNR at the end node $N_i$ can be expressed in terms of MSUE at the relay (see *Appendix 2* for details):

$$GSNR_i = \frac{\beta'^2 |h_{3i}|^2 |h_{i'3}|^2}{\beta'^2 |h_{3i}|^2 MSUE + 1} \quad i = 1,2 \text{ and } i' = 3-i \quad (25)$$

where $\beta' = 2/\sqrt{2(|h_{13}|^2 + |h_{23}|^2) + MSUE}$.

**MMSE-based PNCI**

For general MMSE-based PNCI in (13), the MSUE at the relay node is

$$MSUE_{PNCI\_MMSE} = 2(|h_{13}|^2 + |h_{23}|^2)(2\text{Re}(\lambda)+1) \\ - |\lambda^*+1|^2 E_{y_3}[|\varepsilon(h_{13}x_1 + h_{23}x_2 | y_3)|^2] \quad (26)$$

where $\lambda^* = \dfrac{2(|h_{13}|^2 + |h_{23}|^2)}{E[(h_{13}x_1 + h_{23}x_2)^* E(h_{13}x_1 + h_{23}x_2 \mid y_3)]} - 1$. The derivation of (26), omitted here, is similar to the proof of *Theorem 1* in [9]. The GSNR at the end nodes in terms of MSUE in (26) is of the same form as (25).

We now compare the GSNR performance of PNCF and PNCI. *Theorem 1* in [9] shows that a scaled version of the MMSE estimation is optimal in term of minimizing the MSUE at both relay and the destination. For the PNCF and PNCI schemes, we have the following similar conclusions.

***Theorem 1:*** *In our system, the MMSE based PNCF scheme minimizes the MSUE (Maximizes the GSNR) at the relay node and the two destination nodes among all possible PNCF schemes.*

***Theorem 2:*** *In our system, the MMSE based PNCI scheme minimizes the MSUE (Maximizes the GSNR) at the relay node and the two destination nodes among all possible PNCI schemes.*

The basic idea of the proof of *Theorem 1* is the same as that of *Theorem 2*. It consists of the following two steps. The first step is to prove that MMSE schemes minimize the MSUE at the relay node, which is similar to the proof in [9, *Theorem 1*] and is omitted here. The second step is to prove the equivalence between minimizing MSUE at the relay and maximizing GSNR at the destinations, which can be found in *Appendices 1* and *2*.

### C. Simulation Result

This subsection presents simulation results of the four PNC schemes within the system defined in Section II. Limited by space, we only present results related to the case in which the channels for nodes $N_1$ and $N_2$ are symmetric, i.e. $|h_{13}| = |h_{23}|, |h_{31}| = |h_{32}|$.

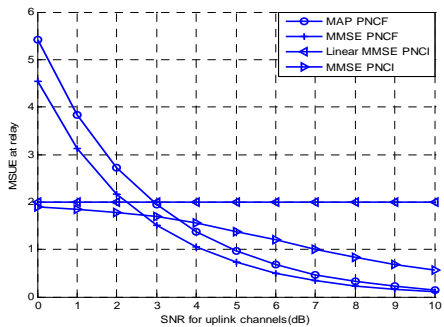

Figure.2 MSUE at the relay when downlink channels are 5dB

We first compare the performance of MMSE and non-MMSE estimations. From Fig.2 and Fig.3, we can see that MMSE estimation is better in terms of maximizing the GSNR and minimizing the MSUE, for each PNC class (PNCI and PNCF). The BER performance is shown in Fig. 3(b). Generally speaking, as in [9] for one-way relay, there is a correspondence between the BER performance and the GSNR performance: a larger GSNR usually leads to a smaller BER with the QPSK modulation in our system.

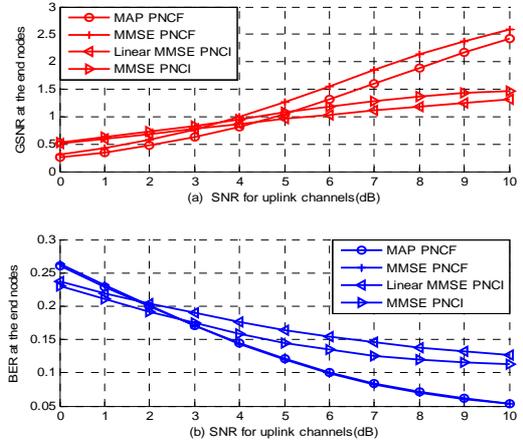

Figure.3 GSNR and BER at end nodes when downlink channels are 5dB

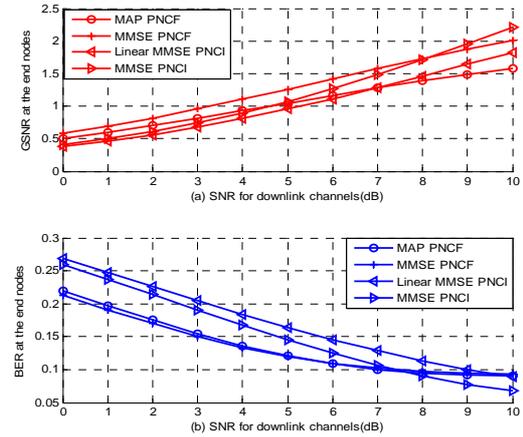

Figure.4 GSNR and BER at the end nodes where uplink channels are 5dB

Let us now compare the performance between the two PNC classes. From the simulation results, we can see that the relative performance of the two classes depends on the quality of the uplink and downlink channels. In particular, Fig. 2 and Fig. 3 show that the PNCF (PNCI) schemes are better when the uplink channel is good (bad). This could be understood intuitively by considering an extreme situation. When the uplink channel is very good, the estimation errors for both PNCI and PNCF schemes are negligible. However, the power needed to send the PNCF signals, $x_1 \oplus x_2$, is 2, while the power needed to send the PNCI signals, $h_{13}x_1 + h_{23}x_2$, is much bigger. Fig. 4 shows that PNCF (PNCI) schemes perform better when the downlink channel is bad (good). This could also be understood by considering another extreme situation. When the downlink channel is so good that all the information available at the relay node can be sent to the end nodes without any loss, PNCF is worse since it loses some information through transforming $h_{13}x_1 + h_{23}x_2$ to $x_1 \oplus x_2$.

## V. CONCLUSION

This paper has proposed a definition for physical-layer network coding (PNC) for two-way relay channels. Specifically, in PNC, a relay node does not decode the information it receives from the two ends separately, but rather will map the combined signals received simultaneously from two ends into a network-coded symbol. Different mapping functions give rise to different possible schemes within the general PNC class. If a particular PNC scheme uses a network code over a finite field, then it belongs to the subclass of PNCF; whereas if it uses a network code over an infinite field, then it belongs to the subclass of PNCI.

With noise, there is an additional issue of estimating the network-coded symbols at the relay. For a given network code, different estimation techniques will give rise to different schemes. We have studied two particular estimation schemes for each of PNCF and PNCI. Our results show that the MMSE schemes outperform the non-MMSE schemes. In addition, when the uplink channel is good and the downlink channel is bad, the PNCF schemes perform better than the PNCI schemes, and vice versa.

This paper has focused on memoryless relay protocols in which channel coding is not applied at the relay node. For PNCF, the end nodes could apply channel codes, such as LDPC code, Turbo code, that are linear for addition over finite field GF(2), i.e. $\oplus$. The relay node would then decode the symbol $x_1 \oplus x_2$ from the combined signals received [12]. For PNCI, however, there is no channel code such that decoding can be performed over the symbol $h_{13}x_1 + h_{23}x_2$.

*Appendix 1*

For PNCF, the GSNR at destination $N_i$ is

$$GSNR_i = \frac{2\beta^2 |h_{3i}|^2}{\beta^2 |h_{3i}|^2 MSUE + 1} \quad i = 1, 2$$

where $\beta = 1/\sqrt{2 + MSUE}$.

**Proof:** Consider $N_1$. For an estimator at the relay, $\hat{X}(r) = x_1 \oplus x_2 + e_u$ with $MSUE = E[|e_u|^2]$, the received signal at $N_1$ is

$$y_1 = h_{31}\beta(x_1 \oplus x_2 + e_u) + n_1$$

By adding its own "self" symbol $x_1$ on the received signal, $N_1$ can obtain

$$y_1' = y_1 \oplus x_1 = h_{31}\beta x_2 + (h_{31}\beta e_u + n_1) \oplus x_1$$

Obviously the information $x_2$ is uncorrelated to $(h_{31}\beta e_u + n_1) \oplus x_1$. Then the GSNR of $y_1'$ can be written as

$$GSNR_1 = \frac{2\beta^2 |h_{31}|^2}{\beta^2 |h_{31}|^2 MSUE + 1} = \frac{2|h_{31}|^2}{(|h_{31}|^2 + 1)MSUE + 2}$$

The last equation shows that maximizing GSNR at the destination is equivalent to minimizing the MSUE at the relay for PNCF schemes. The calculation of $GSNR_2$ is similar.

*Appendix 2:*

For PNCI schemes, the GSNR at destination $N_i$ is

$$GSNR_i = \frac{\beta'^2 |h_{3i}|^2 |h_{i'3}|^2}{\beta'^2 |h_{3i}|^2 MSUE + 1} \quad i = 1,2 \text{ and } i' = 3 - i$$

where $\beta' = 1/\sqrt{2(|h_{13}|^2 + |h_{23}|^2) + MSUE}$.

**Proof:** Consider $N_1$. For an estimator at the relay, $\hat{X}(r) = h_{13}x_1 + h_{23}x_2 + e_u$ with $MSUE = E[|e_u|^2]$, the received signal at $N_1$ is

$$y_1 = h_{31}\beta'(h_{13}x_1 + h_{23}x_2 + e_u) + n_1$$

Making use of the self information $x_1$, $N_1$ can obtain

$$y_1' = y_1 - h_{31}\beta'h_{13}x_1 = h_{31}\beta'h_{23}x_2 + h_{31}\beta'e_u + n_1$$

Since $x_2$ is uncorrelated to $h_{31}\beta'e_u + n_1$, GSNR of $y_1'$ is

$$GSNR_1 = \frac{2\beta'^2 |h_{31}|^2 |h_{23}|^2}{\beta'^2 |h_{31}|^2 MSUE + 1} = \frac{2|h_{31}|^2 |h_{23}|^2}{(|h_{31}|^2 + 1)MSUE + 2(|h_{13}|^2 + |h_{23}|^2)}$$

As in PNCF, the last equation shows that maximizing GSNR at the destination is equivalent to minimizing the MSUE at the relay for PNCI. The calculation of $GSNR_2$ is similar.